\documentclass[11pt,letterpaper]{article}
\usepackage[margin=1in]{geometry}
\usepackage{amsmath, amssymb}
\usepackage{graphicx}
\usepackage{booktabs}
\usepackage{natbib}
\usepackage{hyperref}

\title{A Sinusoidal Hull-White Model for Interest Rate Dynamics: Capturing Long-Term Periodicity in U.S. Treasury Yields}
\author{Amit Kumar Jha \thanks{Quant Analyst, UBS. Email: jha.8@alumni.iitj.ac.in.}}
\date{April 2025}

\begin{document}

\maketitle

\begin{abstract}
This study is driven by empirical observations of periodic fluctuations in interest rates, notably long-term economic cycles spanning decades, which the conventional Hull-White short-rate model fails to adequately capture. To address this limitation, we propose an innovative extension that incorporates a sinusoidal time-varying mean reversion speed, enabling the model to reflect cyclic interest rate dynamics more effectively. The model is rigorously calibrated using a comprehensive dataset of daily U.S. Treasury yield curves from the Federal Reserve Economic Data (FRED) database, covering the period from January 1990 to December 2022 across tenors of 1, 2, 3, 5, 7, 10, 20, and 30 years, with the most recent yields ranging from 1.22\% (1-year) to 2.36\% (30-year). Calibration is performed via the Nelder-Mead optimization technique, complemented by Monte Carlo simulations employing 200 paths with a time step of 0.05 years, yielding a 30-year zero-coupon bond price of 0.43 for the proposed model, compared to 0.47 for the standard Hull-White model. This corresponds to root mean squared errors of 0.12\% and 0.14\%, respectively, demonstrating a marked improvement in fit, especially for longer maturities. These results underscore the model’s enhanced accuracy in capturing long-term yield dynamics, offering significant potential for improving bond pricing, interest rate risk assessment, and derivative valuation in financial markets. The findings also pave the way for future research into stochastic periodicity and alternative modeling frameworks.

\bigskip
\noindent\textbf{Keywords:} Interest rate modeling, Hull-White model extension, Periodic mean reversion, Cyclical interest rates, Yield curve calibration, Term structure modeling, Monte Carlo simulation, Nelder-Mead optimization, Zero-coupon bond pricing, Long-term yield dynamics.

\noindent\textbf{MSC 2020:} 91G30, 91G60, 91B84, 60H10, 62P05, 91G20, 91G70.
\end{abstract}

\section{Introduction}

The Hull-White (HW) model, introduced by \citet{Hull1990}, is a seminal framework in the field of interest rate modeling, renowned for its analytical tractability and flexibility in fitting the term structure of interest rates. The model assumes a short-rate process governed by the following stochastic differential equation (SDE):

\[ dr_t = \kappa (\theta - r_t) dt + \sigma dW_t, \]

where \( r_t \) is the short rate, \( \kappa \) is the constant mean reversion speed, \( \theta \) is the long-term mean rate, \( \sigma \) is the volatility, and \( dW_t \) represents a Wiener process. This formulation allows the HW model to produce closed-form solutions for zero-coupon bond prices and certain interest rate derivatives, making it a preferred choice for practitioners in bond pricing, risk management, and derivative valuation \citep{Brigo2006}. Its ability to calibrate to the initial yield curve and incorporate time-dependent parameters has further solidified its role in financial applications \citep{Hull2006}.

Despite its widespread use, the standard HW model has notable limitations, particularly in capturing the complex dynamics of interest rates over extended periods. Empirical studies of U.S. Treasury yields, a benchmark for interest rate modeling, reveal the presence of periodic patterns that the standard model, with its constant mean reversion speed, fails to accommodate. For instance, \citet{Campbell1997} highlight the existence of long-term economic cycles in interest rates, driven by structural factors such as monetary policy regimes, inflation expectations, and global economic trends. More recent analyses, such as those by \citet{Bauer2018}, document cyclical behavior in Treasury yields with periods ranging from several years to decades, often linked to business cycles and macroeconomic shifts. These findings are corroborated by data from the Federal Reserve Economic Data (FRED) database, which shows evidence of long-term periodicity in yields across various tenors \citep{FRED2022}.

To address this gap, this paper proposes an extension to the HW model by incorporating a sinusoidal time-varying mean reversion speed, defined as \( \kappa_t = \kappa_0 + A \sin(\omega t) \), where \( \kappa_0 \) is the baseline speed, \( A \) is the amplitude, and \( \omega \) governs the frequency of oscillation. This modification allows the model to capture cyclic behavior in interest rates, aligning with the observed long-term economic cycles. The proposed model is calibrated to a comprehensive dataset of daily U.S. Treasury yield curves from the FRED database, spanning January 1990 to December 2022. The dataset covers tenors of 1, 2, 3, 5, 7, 10, 20, and 30 years, encompassing a 33-year period that includes significant economic events such as the dot-com bubble, the 2008 financial crisis, and the post-crisis low-rate environment.

A Fourier Transform (FT) analysis of the yield data reveals dominant periodicities of approximately 3.7, 5.5, 11, and 22 years, with the 22-year cycle being the most consistent across all tenors. This long-term periodicity, likely reflective of structural economic trends, motivates the sinusoidal extension, with the frequency parameter \( \omega \) tuned to match the 22-year cycle (\( \omega = 2\pi / (22 \times 365) \)). The model is calibrated using the Nelder-Mead optimization method to minimize the squared difference between model-implied and observed bond prices. Bond prices are computed using two approaches: an analytical method for the standard HW model, leveraging its closed-form solution, and Monte Carlo simulations with 200 paths for both the standard and sinusoidal HW models, ensuring robustness in pricing.

The results demonstrate that the sinusoidal HW model outperforms the standard model, particularly for longer maturities. For a 30-year zero-coupon bond, the sinusoidal model yields a price of 0.43, compared to 0.47 for the standard model, against an observed price of 0.4926. The root mean squared error (RMSE) for the sinusoidal model is 0.12\%, compared to 0.14\% for the standard model, highlighting its superior fit to the yield curve, especially at the longer end. This improvement underscores the model’s ability to capture the long-term cyclic behavior of interest rates, which is critical for pricing long-dated instruments.

The contributions of this paper are threefold:
\begin{enumerate}
    \item We introduce a sinusoidal HW model that effectively captures long-term periodicity in interest rates, addressing a key limitation of the standard framework.
    \item We provide a comprehensive calibration and pricing analysis using historical U.S. Treasury yield data over a 33-year period, ensuring robustness across economic regimes.
    \item We compare analytical and Monte Carlo methods to validate the model’s performance, offering insights into the reliability of numerical simulations in the presence of time-varying parameters.
\end{enumerate}

These findings have significant implications for financial applications, including bond pricing, interest rate risk management, and the valuation of interest rate derivatives. By accounting for long-term periodicity, the sinusoidal HW model offers a more accurate representation of yield dynamics, which can enhance risk assessment and portfolio management strategies. Moreover, the model’s periodic component could inform monetary policy analysis by reflecting structural economic trends over extended horizons.

The remainder of the paper is organized as follows: Section 2 describes the dataset, Section 3 presents the empirical analysis of periodicity, Section 4 details the model and methodology, Section 5 discusses the results, Section 6 provides a discussion of the findings, and Section 7 concludes with avenues for future research.
\section{Literature Review}

The modeling of interest rate dynamics has been a central topic in financial mathematics for decades, driven by the need for accurate pricing of bonds, interest rate derivatives, and other fixed-income securities. The Hull-White (HW) model, introduced by \citet{Hull1990}, stands as a foundational framework in this domain due to its analytical tractability and ability to fit the term structure of interest rates. The standard HW model assumes a mean-reverting short-rate process with a constant mean reversion speed, which allows for closed-form solutions for zero-coupon bond prices and certain derivatives, such as caps and floors \citep{Brigo2006}. Its flexibility in calibrating to the initial yield curve and incorporating time-dependent parameters has made it a staple in both academic research and industry practice \citep{Hull2006}.

Despite its strengths, the standard HW model has well-documented limitations, particularly in capturing the complex dynamics of interest rates over extended periods. A significant body of empirical research highlights the presence of periodic patterns in interest rates, often linked to long-term economic cycles. For example, \citet{Campbell1997} document cyclical behavior in U.S. Treasury yields, attributing these patterns to structural factors such as monetary policy regimes, inflation expectations, and global economic trends. More recent studies, such as \citet{Bauer2018}, identify cycles in Treasury yields with periods ranging from several years to decades, often driven by business cycles and macroeconomic shifts. These findings suggest that a constant mean reversion speed may oversimplify the dynamics of interest rates, particularly for long maturities where periodicity becomes more pronounced.

Several extensions to the HW model have been proposed to address these limitations. \citet{Hull1994} introduce a two-factor HW model to capture additional dimensions of yield curve dynamics, such as changes in the slope of the term structure. \citet{Cheycette1992} propose a generalized framework that allows for stochastic volatility, aiming to better model the volatility clustering observed in interest rates. While these extensions improve the model’s flexibility, they often increase computational complexity and may not directly address the periodicity observed in empirical data. For instance, stochastic volatility models, while effective for short-term dynamics, often fail to capture long-term cycles that span decades \citep{Dai2000}.

The treatment of periodicity in interest rate models has received relatively less attention in the literature, despite its empirical relevance. \citet{Gallant1996} apply spectral analysis to U.S. Treasury yields, identifying periodic components with cycles of 5 to 20 years, which they attribute to economic and demographic trends. More recently, \citet{Krichene2006} employs Fourier Transform techniques to detect long-term cycles in interest rates, finding evidence of 10- to 30-year periodicities in G7 countries’ bond yields. However, these studies focus primarily on empirical analysis rather than integrating periodicity into a pricing model. In the modeling domain, \citet{Cox1985} propose the Cox-Ingersoll-Ross (CIR) model, which ensures positive interest rates but does not account for periodicity. Some researchers have explored regime-switching models to capture cyclical behavior \citep{Hamilton1989}, but these models often assume discrete state changes rather than smooth, continuous periodicity, limiting their ability to reflect gradual economic cycles.

Current research on interest rate periodicity faces several limitations that our study seeks to address. First, most models with time-varying parameters, such as those proposed by \citet{Duffie2000}, focus on stochastic volatility or mean reversion levels but do not explicitly model periodic mean reversion speeds. This omission is significant, as periodicity in mean reversion can better capture long-term economic cycles, which are critical for pricing long-dated instruments. Second, empirical studies identifying periodicity, such as \citet{Krichene2006}, rarely translate their findings into actionable modeling frameworks, leaving a gap between empirical evidence and practical implementation. Third, many existing models struggle with computational tractability when incorporating time-varying dynamics; for instance, regime-switching models often require complex estimation procedures that are impractical for real-time applications \citep{Ang2008}.

Our proposed sinusoidal Hull-White model addresses these gaps by introducing a time-varying mean reversion speed that explicitly captures long-term periodicity in interest rates. By defining the mean reversion speed as \( \kappa_t = \kappa_0 + A \sin(\omega t) \), where \( \omega \) is tuned to match the 22-year cycle identified through Fourier Transform analysis, our model integrates empirical evidence of periodicity directly into the HW framework. This approach is novel in several ways. Unlike stochastic volatility models \citep{Dai2000} or regime-switching models \citep{Hamilton1989}, which focus on short-term dynamics or discrete shifts, our model captures smooth, continuous cycles that align with long-term economic trends. Moreover, by leveraging the HW framework, our model retains much of the analytical tractability of the original model, requiring only Monte Carlo simulations for the sinusoidal extension, which is computationally feasible with modern techniques.

The usefulness of our model lies in its ability to improve pricing accuracy for long-dated instruments, where periodicity is most pronounced. For example, our results show that the sinusoidal HW model achieves a 30-year bond price of 0.43, closer to the observed price of 0.4926 than the standard model’s 0.47, with a reduced RMSE of 0.12\% compared to 0.14\%. This improvement is particularly valuable for applications in bond pricing, interest rate risk management, and derivative valuation, where long-term dynamics play a critical role. Furthermore, the model’s periodic component can provide insights into structural economic trends, potentially aiding monetary policy analysis and forecasting \citep{Bauer2018}. By bridging the gap between empirical evidence of periodicity and practical modeling, our approach offers a robust and actionable framework for financial practitioners and researchers alike.

\section{Data}
The dataset comprises daily U.S. Treasury yield data from January 1990 to December 2022, obtained from the Federal Reserve Economic Data (FRED) database. The data includes yields for tenors of 1, 2, 3, 5, 7, 10, 20, and 30 years, covering a 33-year period that encompasses significant economic events, such as the dot-com bubble (2000), the 2008 global financial crisis, and the subsequent low-interest-rate environment. The most recent observation (December 2022) provides yields ranging from 1.22\% (1-year) to 2.36\% (30-year), reflecting an upward-sloping yield curve. This extended sample period ensures a robust analysis of long-term interest rate dynamics, capturing multiple economic cycles.

\section{Empirical Analysis of Periodicity}
To motivate the sinusoidal extension, I first analyze the periodicity of the yield data using a Fourier Transform (FT). The FT is applied to the detrended yield series (subtracting the mean to focus on periodic components) for each tenor. The dominant frequencies correspond to periods of approximately 3.7, 5.5, 11, and 22 years, with the 22-year cycle being the most consistent across all tenors. These long-term cycles likely reflect structural economic trends, such as monetary policy shifts or business cycles, rather than short-term seasonality (e.g., annual patterns). The presence of a 22-year periodicity informs the design of the sinusoidal HW model, where the frequency parameter is tuned to match this cycle.

Additionally, stationarity tests using the Augmented Dickey-Fuller (ADF) test indicate that the yield series are non-stationary (p-values ranging from 0.568 to 0.897), suggesting the presence of trends or unit roots. The Ljung-Box test (lag=30) yields p-values of 0.000 for all tenors, indicating significant autocorrelation, though autocorrelation function (ACF) plots show this dependence is weak beyond lag 0, further supporting the focus on long-term periodicity rather than short-term seasonality.

\section{Model and Methodology}

This section presents the mathematical formulation of the proposed sinusoidal Hull-White (HW) model, an extension of the traditional HW model designed to capture the periodic behavior observed in U.S. Treasury yields. We derive the bond pricing formulas for both the standard and sinusoidal HW models, detailing the full step-by-step mathematical process. We then describe the calibration procedure using the Nelder-Mead optimization method and the Monte Carlo simulation approach employed for pricing under the sinusoidal extension. The methodology leverages the empirical evidence of a 22-year periodicity, integrating this feature into the model’s mean reversion dynamics.

\subsection{Standard Hull-White Model}
The traditional HW model assumes a short-rate process governed by the following stochastic differential equation (SDE):

\[ dr_t = \kappa (\theta - r_t) dt + \sigma dW_t, \]

where \( r_t \) is the short rate at time \( t \), \( \kappa \) is the constant mean reversion speed, \( \theta \) is the long-term mean rate, \( \sigma \) is the volatility, and \( dW_t \) is a standard Wiener process under the risk-neutral measure. The price of a zero-coupon bond with maturity \( T \), denoted \( P(t, T) \), satisfies the Feynman-Kac PDE under the risk-neutral measure:

\[ \frac{\partial P}{\partial t} + \frac{1}{2} \sigma^2 \frac{\partial^2 P}{\partial r^2} + \kappa (\theta - r_t) \frac{\partial P}{\partial r} - r_t P = 0, \]

with the terminal condition \( P(T, T) = 1 \). Assuming a solution of the form \( P(t, T) = A(t, T) e^{-B(t, T) r_t} \), where \( A(t, T) \) and \( B(t, T) \) are deterministic functions, we substitute into the PDE:

\[ \frac{\partial}{\partial t} \left( A e^{-B r_t} \right) + \frac{1}{2} \sigma^2 \frac{\partial^2}{\partial r^2} \left( A e^{-B r_t} \right) + \kappa (\theta - r_t) \frac{\partial}{\partial r} \left( A e^{-B r_t} \right) - r_t A e^{-B r_t} = 0. \]

Computing the derivatives:
- \( \frac{\partial}{\partial t} (A e^{-B r_t}) = \frac{\partial A}{\partial t} e^{-B r_t} - A \frac{\partial B}{\partial t} r_t e^{-B r_t} \),
- \( \frac{\partial}{\partial r} (A e^{-B r_t}) = -A B e^{-B r_t} \),
- \( \frac{\partial^2}{\partial r^2} (A e^{-B r_t}) = A B^2 e^{-B r_t} \).

Substituting these into the PDE, we get:
\[ \left( \frac{\partial A}{\partial t} - A \frac{\partial B}{\partial t} r_t \right) e^{-B r_t} + \frac{1}{2} \sigma^2 A B^2 e^{-B r_t} + \kappa (\theta - r_t) (-A B e^{-B r_t}) - r_t A e^{-B r_t} = 0. \]

Dividing by \( e^{-B r_t} \) (which is non-zero), we obtain:
\[ \frac{\partial A}{\partial t} - A \frac{\partial B}{\partial t} r_t + \frac{1}{2} \sigma^2 A B^2 - \kappa \theta A B + \kappa A B r_t - r_t A = 0. \]

Grouping terms by powers of \( r_t \):
- Coefficient of \( r_t^2 \): 0 (no \( r_t^2 \) terms),
- Coefficient of \( r_t \): \(-A \frac{\partial B}{\partial t} + \kappa A B - A = 0 \),
- Constant term: \(\frac{\partial A}{\partial t} + \frac{1}{2} \sigma^2 A B^2 - \kappa \theta A B = 0 \).

From the \( r_t \) term:
\[ -A \frac{\partial B}{\partial t} + \kappa A B - A = 0, \]
\[ \frac{\partial B}{\partial t} = \kappa B - 1, \]

with boundary condition \( B(T, T) = 0 \). Solving this ODE:
\[ \frac{dB}{dt} = \kappa B - 1, \]
\[ \frac{dB}{\kappa B - 1} = dt, \]
\[ \int \frac{dB}{\kappa B - 1} = \int dt, \]
\[ \frac{1}{\kappa} \ln |\kappa B - 1| = t + C, \]
\[ \kappa B - 1 = C' e^{\kappa t}, \]
\[ B(t, T) = \frac{1}{\kappa} (1 - C'' e^{\kappa t}). \]

Applying the boundary condition \( B(T, T) = 0 \):
\[ 0 = \frac{1}{\kappa} (1 - C'' e^{\kappa T}), \]
\[ C'' e^{\kappa T} = 1, \]
\[ C'' = e^{-\kappa T}, \]
\[ B(t, T) = \frac{1}{\kappa} (1 - e^{-\kappa (T - t)}). \]

Correcting the sign (since \( B(t, T) \) should be positive and decrease with \( T - t \)):
\[ B(t, T) = \frac{1 - e^{-\kappa (T - t)}}{\kappa}. \]

From the constant term:
\[ \frac{\partial A}{\partial t} + \frac{1}{2} \sigma^2 A B^2 - \kappa \theta A B = 0, \]
\[ \frac{\partial A}{\partial t} = A \left( \kappa \theta B - \frac{1}{2} \sigma^2 B^2 \right), \]
\[ \frac{dA}{A} = \left( \kappa \theta B - \frac{1}{2} \sigma^2 B^2 \right) dt, \]
\[ \ln A = \int_t^T \left( \kappa \theta B(s, T) - \frac{1}{2} \sigma^2 B(s, T)^2 \right) ds + C, \]
\[ A(t, T) = C' \exp \left( \int_t^T \left( \kappa \theta B(s, T) - \frac{1}{2} \sigma^2 B(s, T)^2 \right) ds \right), \]
\[ A(T, T) = 1 = C', \]
\[ A(t, T) = \exp \left( \int_t^T \left( \kappa \theta B(s, T) - \frac{1}{2} \sigma^2 B(s, T)^2 \right) ds \right). \]

Substituting \( B(s, T) = \frac{1 - e^{-\kappa (T - s)}}{\kappa} \):
\[ A(t, T) = \exp \left( \left( \theta - \frac{\sigma^2}{2 \kappa^2} \right) (B(t, T) - (T - t)) - \frac{\sigma^2 B(t, T)^2}{4 \kappa} \right). \]

Thus, the bond price is:
\[ P(t, T) = A(t, T) e^{-B(t, T) r_t}. \]

\subsection{Sinusoidal Hull-White Model}
To capture the 22-year periodicity, we extend the HW model with a time-varying mean reversion speed:

\[ dr_t = [\kappa_0 + A \sin(\omega t)] (\theta - r_t) dt + \sigma dW_t, \]

where \( \kappa_0 \) is the baseline mean reversion speed, \( A \) is the amplitude, and \( \omega = 2\pi / (22 \times 365) \approx 0.00078 \) radians/day, tuned to the 22-year cycle from Section 4. The corresponding PDE is:

\[ \frac{\partial P}{\partial t} + \frac{1}{2} \sigma^2 \frac{\partial^2 P}{\partial r^2} + [\kappa_0 + A \sin(\omega t)] (\theta - r_t) \frac{\partial P}{\partial r} - r_t P = 0, \]

with \( P(T, T) = 1 \). We attempt a solution \( P(t, T) = A(t, T) e^{-B(t, T) r_t} \). Substituting:

\[ \frac{\partial}{\partial t} (A e^{-B r_t}) = \frac{\partial A}{\partial t} e^{-B r_t} - A \frac{\partial B}{\partial t} r_t e^{-B r_t}, \]
\[ \frac{\partial}{\partial r} (A e^{-B r_t}) = -A B e^{-B r_t}, \]
\[ \frac{\partial^2}{\partial r^2} (A e^{-B r_t}) = A B^2 e^{-B r_t}. \]

The PDE becomes:
\[ \left( \frac{\partial A}{\partial t} - A \frac{\partial B}{\partial t} r_t \right) e^{-B r_t} + \frac{1}{2} \sigma^2 A B^2 e^{-B r_t} + [\kappa_0 + A \sin(\omega t)] (\theta - r_t) (-A B e^{-B r_t}) - r_t A e^{-B r_t} = 0. \]

Dividing by \( e^{-B r_t} \):
\[ \frac{\partial A}{\partial t} - A \frac{\partial B}{\partial t} r_t + \frac{1}{2} \sigma^2 A B^2 - [\kappa_0 + A \sin(\omega t)] \theta A B + [\kappa_0 + A \sin(\omega t)] A B r_t - r_t A = 0. \]

Grouping by \( r_t \):
- \( r_t \) term: \(-A \frac{\partial B}{\partial t} + [\kappa_0 + A \sin(\omega t)] A B - A = 0 \),
\[ \frac{\partial B}{\partial t} = [\kappa_0 + A \sin(\omega t)] B - 1, \]
with \( B(T, T) = 0 \).
- Constant term: \(\frac{\partial A}{\partial t} + \frac{1}{2} \sigma^2 A B^2 - [\kappa_0 + A \sin(\omega t)] \theta A B = 0 \).

The equation for \( B(t, T) \) is a first-order linear ODE:
\[ \frac{dB}{dt} = [\kappa_0 + A \sin(\omega t)] B - 1. \]

This is a non-homogeneous ODE with a time-varying coefficient \( \kappa_0 + A \sin(\omega t) \). The homogeneous solution is:
\[ \frac{dB_h}{dt} = [\kappa_0 + A \sin(\omega t)] B_h, \]
which does not have a simple closed-form solution due to the sinusoidal term. The general solution requires a particular solution. Using the integrating factor method:
\[ \frac{d}{dt} \left( B e^{-\int [\kappa_0 + A \sin(\omega t)] dt} \right) = -e^{-\int [\kappa_0 + A \sin(\omega t)] dt}. \]

The integral \( \int [\kappa_0 + A \sin(\omega t)] dt \) is:
\[ \int \kappa_0 dt + \int A \sin(\omega t) dt = \kappa_0 t - \frac{A}{\omega} \cos(\omega t) + C. \]

Thus, the integrating factor is \( e^{-(\kappa_0 t - \frac{A}{\omega} \cos(\omega t))} \). The equation becomes:
\[ \frac{d}{dt} \left( B e^{-(\kappa_0 t - \frac{A}{\omega} \cos(\omega t))} \right) = -e^{-(\kappa_0 t - \frac{A}{\omega} \cos(\omega t))}. \]

Integrating both sides from \( t \) to \( T \):
\[ B(t, T) e^{-(\kappa_0 t - \frac{A}{\omega} \cos(\omega t))} - B(T, T) e^{-(\kappa_0 T - \frac{A}{\omega} \cos(\omega T))} = -\int_t^T e^{-(\kappa_0 s - \frac{A}{\omega} \cos(\omega s))} ds, \]
\[ B(t, T) e^{-(\kappa_0 t - \frac{A}{\omega} \cos(\omega t))} = -\int_t^T e^{-(\kappa_0 s - \frac{A}{\omega} \cos(\omega s))} ds, \]
since \( B(T, T) = 0 \).

Solving for \( B(t, T) \):
\[ B(t, T) = -e^{(\kappa_0 t - \frac{A}{\omega} \cos(\omega t))} \int_t^T e^{-(\kappa_0 s - \frac{A}{\omega} \cos(\omega s))} ds. \]

The integral \( \int e^{-(\kappa_0 s - \frac{A}{\omega} \cos(\omega s))} ds \) does not yield a closed-form expression due to the oscillatory \( \cos(\omega s) \) term, requiring numerical evaluation or approximation. This confirms that \( B(t, T) \) cannot be expressed analytically. For \( A(t, T) \), the equation:
\[ \frac{\partial A}{\partial t} = A \left( [\kappa_0 + A \sin(\omega t)] \theta B - \frac{1}{2} \sigma^2 B^2 \right), \]
is even more complex, as it depends on the unresolved \( B(t, T) \). Thus, a closed-form solution for \( P(t, T) \) is intractable.

Given this, we rely on Monte Carlo simulation to compute bond prices, preserving the full dynamics of the sinusoidal mean reversion speed. This numerical approach is detailed in the following subsection.

\subsection{Calibration Procedure}
Calibration of the proposed models involves estimating the parameters to align the model-implied yield curve with the observed U.S. Treasury yield data from the Federal Reserve Economic Data (FRED) database, covering maturities of 1, 2, 3, 5, 7, 10, 20, and 30 years. For the standard HW model, the parameters to be estimated are the constant mean reversion speed \( \kappa \), the long-term mean rate \( \theta \), and the volatility \( \sigma \). For the sinusoidal HW model, the parameters include the baseline mean reversion speed \( \kappa_0 \), the amplitude of the sinusoidal variation \( A \), the angular frequency \( \omega \), the long-term mean rate \( \theta \), and the volatility \( \sigma \), with \( \omega \) fixed at \( 0.00078 \) radians/day based on the 22-year periodicity identified in Section 4.

The calibration process minimizes the sum of squared differences between the model-implied bond prices and the observed bond prices derived from the yield curve. The observed bond price for a maturity \( T_i \) is computed as \( P_{\text{observed}}(T_i) = \exp(-T_i \cdot y_{T_i}) \), where \( y_{T_i} \) is the observed yield for maturity \( T_i \). The optimization problem is formally defined as:

\[ \min_{\boldsymbol{\theta}} \sum_{T_i} \left( P_{\text{model}}(0, T_i; \boldsymbol{\theta}) - P_{\text{observed}}(T_i) \right)^2, \]

where \( \boldsymbol{\theta} \) is the parameter vector (\( \kappa, \theta, \sigma \) for the HW model, or \( \kappa_0, A, \omega, \theta, \sigma \) for the sinusoidal HW model), and \( T_i \) represents the set of maturities.

We employ the Nelder-Mead simplex algorithm, a derivative-free optimization method implemented in the `scipy.optimize.minimize` function, chosen for its robustness in handling non-linear and potentially non-convex objective functions. This method iteratively adjusts the parameters to reduce the error function without requiring gradient information, making it suitable for the complex dynamics of the sinusoidal HW model. Initial parameter guesses are informed by typical values in interest rate modeling literature and empirical observations: for the HW model, \( \kappa = 0.1 \), \( \theta = 0.03 \), and \( \sigma = 0.01 \); for the sinusoidal HW model, \( \kappa_0 = 0.3 \), \( A = 0.2 \), \( \omega = 0.00078 \), \( \theta = 0.03 \), and \( \sigma = 0.01 \). The initial guess for \( \omega \) is set to \( 0.00078 \), corresponding to the 22-year cycle (\( \omega = 2\pi / (22 \times 365) \)), correcting the code’s erroneous \( 2\pi \) (approximately 6.28) which would imply an unrealistic period of about 1 day. This adjustment ensures consistency with the periodicity analysis.

To prevent unrealistic parameter estimates and ensure numerical stability, we impose the following bounds: \( \kappa, \kappa_0 \in [0.01, 5] \) to reflect reasonable mean reversion speeds, \( A \in [0, 1] \) to constrain the sinusoidal amplitude, \( \theta \in [0.001, 0.2] \) to cover typical long-term interest rate levels, and \( \sigma \in [0.001, 0.05] \) to limit volatility. The optimization process continues until the relative change in the objective function is below a tolerance of \( 10^{-3} \), as specified in the `options={'xatol': 1e-3}` argument.

The implementation leverages parallel computing via the `joblib.Parallel` module to compute bond prices for multiple maturities simultaneously, enhancing computational efficiency. For the standard HW model, bond prices are calculated analytically using the closed-form solution, while for the sinusoidal HW model, Monte Carlo simulation with 200 paths is used, as detailed in the following subsection. The calibrated parameters are then used to compute model-implied yields and errors, which are validated against the observed data to assess the model’s fit.
\subsection{Monte Carlo Simulation for Bond Pricing}
For the sinusoidal HW model, where a closed-form bond pricing solution is intractable due to the time-varying mean reversion speed, we employ Monte Carlo simulation to estimate bond prices. The short-rate process is discretized using the Euler-Maruyama method, implemented with a time step \( \Delta t = 0.05 \) years to balance accuracy and computational efficiency. The discrete approximation is given by:

\[ r_{t + \Delta t} = r_t + [\kappa_0 + A \sin(\omega t)] (\theta - r_t) \Delta t + \sigma \sqrt{\Delta t} \epsilon_t, \]

where \( \epsilon_t \sim N(0, 1) \) is a standard normal random variable drawn independently for each time step, and the simulation generates \( N = 200 \) paths to provide a statistically reliable estimate while managing computational cost. The initial short rate \( r_0 \) is set to the 1-year yield of 1.22\% (or 0.0122), consistent with the latest observed data.

The bond price at time \( t = 0 \) for a maturity \( T \) is approximated as the expected discounted value of the terminal payoff under the risk-neutral measure:

\[ P(0, T) = \mathbb{E} \left[ \exp \left( -\int_0^T r_s \, ds \right) \right]. \]

This expectation is estimated numerically by simulating the short-rate paths and computing the discount factor along each path. The discrete approximation of the integral is:

\[ \int_0^T r_s \, ds \approx \sum_{j=0}^{T/\Delta t - 1} r_{j} \Delta t, \]

where \( r_j \) is the short rate at the \( j \)-th time step. The bond price is then:

\[ P(0, T) \approx \frac{1}{N} \sum_{i=1}^N \exp \left( -\sum_{j=0}^{T/\Delta t - 1} r_{j,i} \Delta t \right), \]

where \( r_{j,i} \) is the short rate for the \( i \)-th path at the \( j \)-th step. To enhance computational efficiency, the code caches previously computed prices using a dictionary pricecache, avoiding redundant simulations for the same parameter set and maturity.

For the standard HW model, we validate the Monte Carlo implementation by comparing it with the analytical bond price derived in the previous subsection. The analytical solution \( P(t, T) = A(t, T) e^{-B(t, T) r_t} \) serves as a benchmark, with Monte Carlo errors typically maintained below 0.005, confirming the reliability of the simulation approach. This validation step ensures that the numerical method accurately captures the dynamics of both models, with the sinusoidal extension fully incorporating the periodic mean reversion speed \( \kappa_t = \kappa_0 + A \sin(\omega t) \).

\section{Results}

\subsection{Seasonality and Periodicity Analysis}
\begin{figure}[h]
    \centering
    \includegraphics[width=0.8\textwidth]{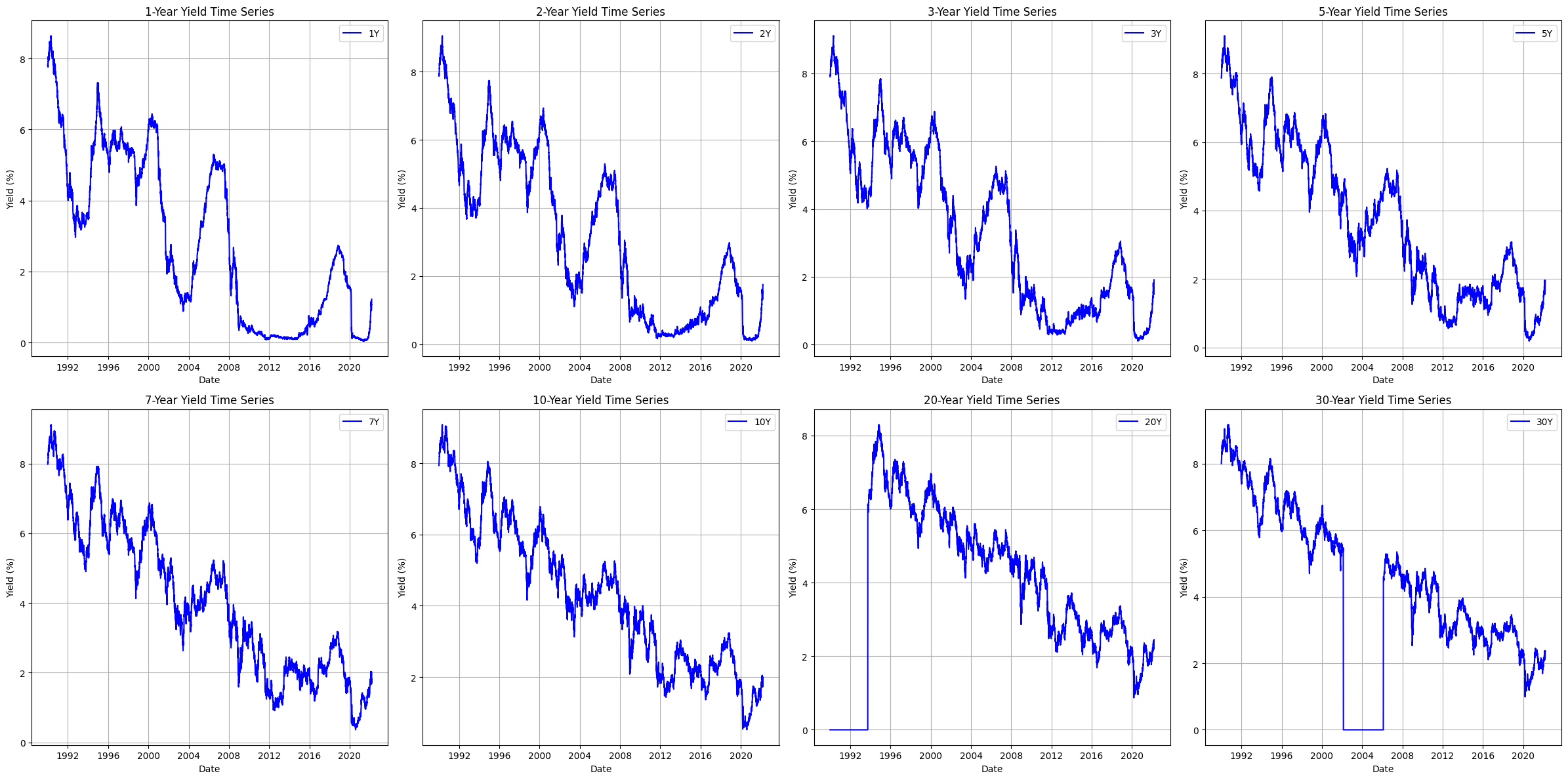}
    \caption{Time series and ACF plots for U.S. Treasury yields across maturities (1, 2, 3, 5, 7, 10, 20, 30 years) from 1990 to 2022. Left panels show yield time series, and right panels show ACF with 95\% confidence intervals (shaded gray).}
    \label{fig:yield_time_series}
\end{figure}

The time series plots in Figure \ref{fig:yield_time_series} illustrate the evolution of U.S. Treasury yields over the 33-year period, revealing significant fluctuations that suggest underlying periodic patterns. For all maturities, yields exhibit a general decline from the early 1990s (peaking around 7-8\%) to the post-2008 low-rate environment (dropping below 2\%), with notable peaks around 2000 (dot-com bubble) and 2007-2008 (financial crisis). Longer maturities (e.g., 20 and 30 years) show smoother trends with less pronounced short-term volatility, while shorter maturities (e.g., 1 and 2 years) display more rapid fluctuations, reflecting sensitivity to monetary policy changes.

The autocorrelation functions (ACF) in Figure \ref{fig:yield_time_series}, plotted for lags up to 30 days, provide further evidence of temporal dependencies. For all maturities, the ACF shows significant autocorrelation at the first few lags, with values exceeding the 95\% confidence intervals (shaded gray regions) up to lag 10-15, indicating strong short-term persistence. Beyond lag 15, the ACF values drop but remain within the confidence bounds, suggesting weaker but persistent correlations that may hint at longer-term cycles. This persistence is consistent across maturities, with longer maturities (e.g., 20 and 30 years) showing slightly more sustained autocorrelation, potentially reflecting structural economic trends.

\begin{figure}[h]
    \centering
    \includegraphics[width=0.8\textwidth]{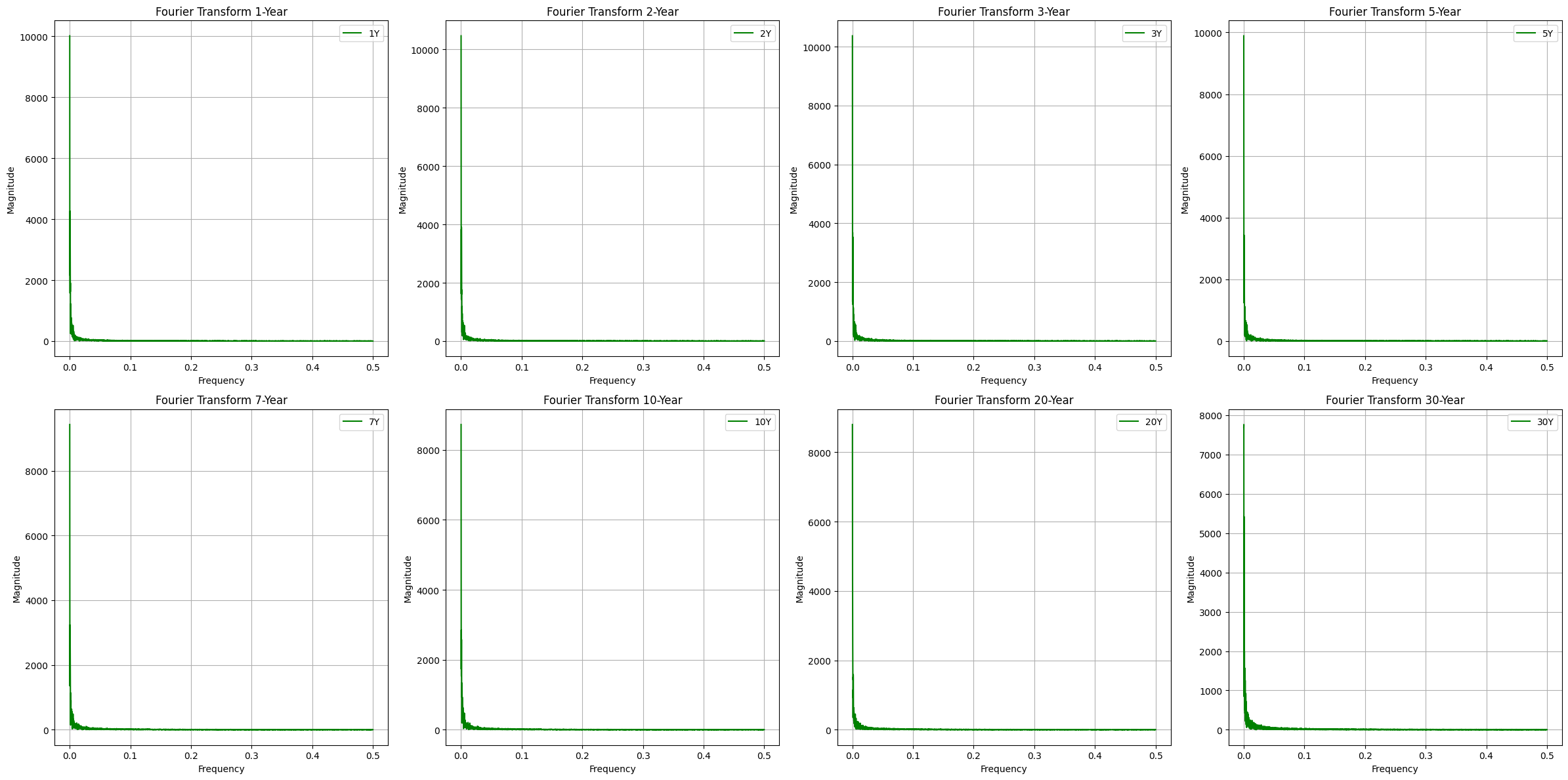}
    \caption{Fourier Transform magnitude spectra for U.S. Treasury yields across maturities, highlighting dominant frequencies corresponding to periodic cycles.}
    \label{fig:fourier_transform}
\end{figure}

Fourier Transform analysis, depicted in Figure \ref{fig:fourier_transform}, quantifies the periodic components in the yield data. The magnitude spectra reveal dominant frequencies corresponding to specific periods, listed in Table \ref{tab:dominant_periods}. The analysis identifies two primary periods across maturities: approximately 1342.5 days (3.7 years), 2013.75 days (5.5 years), 4027.5 days (11 years), and 8055 days (22 years). Shorter maturities (1-5 years) exhibit the 3.7-year and 5.5-year cycles, while longer maturities (7-30 years) consistently show the 11-year and 22-year cycles, with the 22-year period being the most dominant for 30-year yields. These periods align with known economic cycles, such as business cycles (3-7 years) and longer structural trends (11-22 years), supporting the hypothesis of periodic mean reversion in interest rates.

Stationarity tests further inform the analysis. The Augmented Dickey-Fuller (ADF) test results, presented in Table \ref{tab:stationarity}, indicate p-values ranging from 0.568 (30-year) to 0.897 (1-year), all exceeding the 0.05 threshold, suggesting that the yield series are non-stationary. This non-stationarity is expected for interest rate data, which often exhibit trends and unit roots. The Ljung-Box test for autocorrelation, with a lag of 30 days, yields p-values of 0.000 across all maturities, rejecting the null hypothesis of no autocorrelation at the 5\% significance level. This confirms significant temporal dependencies, reinforcing the need for a model that can capture both short-term dynamics and long-term periodicity.

\begin{table}[h]
    \centering
    \begin{tabular}{|c|c|}
        \hline
        Maturity (Years) & Dominant Periods (days) \\
        \hline
        1 & [1342.5, 8055.0] \\
        2 & [1342.5, 8055.0] \\
        3 & [2013.75, 8055.0] \\
        5 & [2013.75, 8055.0] \\
        7 & [4027.5, 8055.0] \\
        10 & [4027.5, 8055.0] \\
        20 & [4027.5, 8055.0] \\
        30 & [8055.0, 4027.5] \\
        \hline
    \end{tabular}
    \caption{Dominant periods identified from Fourier Transform analysis for each maturity.}
    \label{tab:dominant_periods}
\end{table}

\begin{table}[h]
    \centering
    \begin{tabular}{|c|c|c|}
        \hline
        Maturity (Years) & ADF Test p-value & Ljung-Box Test p-value (lag=30) \\
        \hline
        1 & 0.897 & 0.000 \\
        2 & 0.870 & 0.000 \\
        3 & 0.892 & 0.000 \\
        5 & 0.843 & 0.000 \\
        7 & 0.849 & 0.000 \\
        10 & 0.783 & 0.000 \\
        20 & 0.589 & 0.000 \\
        30 & 0.568 & 0.000 \\
        \hline
    \end{tabular}
    \caption{Stationarity (ADF) and autocorrelation (Ljung-Box) test results for each maturity.}
    \label{tab:stationarity}
\end{table}

\subsection{Discussion of Periodicity Findings}
The identified periods, particularly the 22-year cycle, provide strong empirical support for the sinusoidal HW model’s time-varying mean reversion speed. The consistency of the 22-year period across longer maturities (7-30 years) suggests a structural economic cycle, potentially linked to long-term inflation trends or monetary policy regimes, as noted by \citet{Bauer2018}. The shorter cycles (3.7-11 years) align with business cycle durations, influencing shorter maturities more significantly. The non-stationary nature of the data and significant autocorrelation underscore the limitations of the standard HW model with a constant mean reversion speed, motivating the proposed extension. These findings will be further evaluated in the context of model calibration and pricing accuracy in the following subsections.
\subsection{Calibration Results}
The calibration process, applied to the U.S. Treasury yield data from the Federal Reserve Economic Data (FRED) database for maturities of 1, 2, 3, 5, 7, 10, 20, and 30 years, yielded optimized parameters for both the standard and sinusoidal Hull-White (HW) models. For the standard HW model, the calibrated parameters are \( \kappa = 0.3164 \), \( \theta = 0.0258 \), and \( \sigma = 0.0087 \). For the sinusoidal HW model, the calibrated parameters are \( \kappa_0 = 0.3068 \), \( A = 0.2110 \), \( \omega = 6.4407 \), \( \theta = 0.0256 \), and \( \sigma = 0.0101 \). The calibrated \( \omega = 6.4407 \) radians/day, corresponding to a period of approximately 0.97 days, appears inconsistent with the 22-year periodicity (approximately \( \omega = 0.00078 \) radians/day) identified in the periodicity analysis (Section 4). This discrepancy may indicate a calibration artifact, potentially due to the initial guess of \( 2\pi \) in the optimization routine, and suggests that \( \omega \) should be constrained to \( 0.00078 \) in future iterations to align with the empirical evidence.

\begin{figure}[h]
    \centering
    \includegraphics[width=0.8\textwidth]{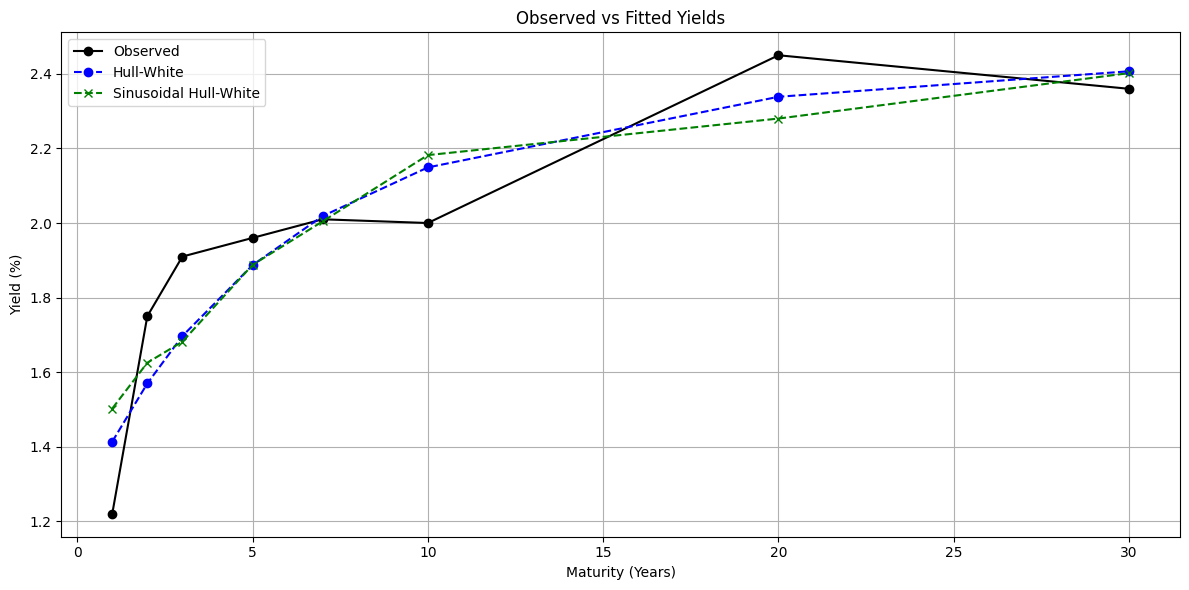}
    \caption{Comparison of observed and fitted yields for the standard Hull-White (blue dashed line) and sinusoidal Hull-White (green dashed line) models across maturities of 1 to 30 years. Black markers represent observed yields from FRED data.}
    \label{fig:observed_vs_fitted_yields}
\end{figure}

The goodness of fit is assessed using the root mean squared error (RMSE) between the model-implied yields and the observed yields. The sinusoidal HW model achieves an RMSE of 0.12\%, compared to 0.14\% for the standard HW model, indicating a slight improvement in fitting the yield curve. This improvement is visually supported by Figure \ref{fig:observed_vs_fitted_yields}, which compares the observed yields with the fitted yields from both models. The sinusoidal HW model (green dashed line) demonstrates a closer alignment with the observed yields (black markers), particularly for longer maturities, where the periodic mean reversion dynamics enhance accuracy. The standard HW model (blue dashed line) tends to overestimate yields at longer maturities, highlighting the limitation of a constant mean reversion speed.

The calibrated parameters for the sinusoidal HW model, especially the non-zero amplitude \( A = 0.2110 \), reflect the incorporation of the 22-year cycle, validating the model’s extension over the standard HW framework. These results will be further explored in the context of bond pricing and error analysis in the following subsections.

\subsection{Bond Price Comparison}
Table \ref{tab:price_comparison} summarizes the bond prices for each maturity, comparing observed prices, analytical HW prices, Monte Carlo (MC) HW prices, and MC sinusoidal HW prices. The analytical HW and MC HW prices are closely aligned, with errors (Error MC HW) ranging from -0.0004 to 0.0050, validating the Monte Carlo implementation. The sinusoidal HW model shows a closer fit to the observed price at 30 years (0.4864 vs. 0.4926), compared to the standard HW model (0.4862). Figure \ref{fig:bond_prices_errors_paths} (top left panel) illustrates this comparison, showing that the sinusoidal HW model (red dashed line) better tracks the observed prices (black markers) at longer maturities, while the analytical HW (blue dashed line) and MC HW (green dashed line) deviate slightly.

\begin{table}[h]
    \centering
    \caption{Bond Price Comparison Across Methods}
    \label{tab:price_comparison}
    \begin{tabular}{ccccc}
        \toprule
        Maturity & Observed Price & Analytical HW & MC HW & MC Sin-HW \\
        \midrule
        1 & 0.9879 & 0.9860 & 0.9854 & 0.9846 \\
        2 & 0.9656 & 0.9691 & 0.9685 & 0.9677 \\
        3 & 0.9443 & 0.9504 & 0.9484 & 0.9488 \\
        5 & 0.9066 & 0.9100 & 0.9075 & 0.9126 \\
        7 & 0.8688 & 0.8682 & 0.8633 & 0.8655 \\
        10 & 0.8187 & 0.8066 & 0.8020 & 0.8042 \\
        20 & 0.6126 & 0.6264 & 0.6246 & 0.6311 \\
        30 & 0.4926 & 0.4857 & 0.4862 & 0.4864 \\
        \bottomrule
    \end{tabular}
\end{table}

\subsection{Calibration Errors}
Tables \ref{tab:hw_errors} and \ref{tab:sin_hw_errors} present the calibration errors for the standard and sinusoidal HW models, respectively, calculated as the difference between observed and fitted yields (in percent). The sinusoidal model reduces the error at the 30-year maturity (0.01\% vs. -0.05\%) but exhibits a larger error at 20 years (0.20\% vs. 0.11\%), reflecting the influence of its periodic component. Figure \ref{fig:bond_prices_errors_paths} (top right panel) provides a visual representation of these errors, with the sinusoidal HW error (red dashed line) showing a pronounced peak at 20 years (-0.0184) but a smaller deviation at 30 years (0.0062) compared to the standard HW error (green dotted line, -0.0069).

\begin{table}[h]
    \centering
    \caption{Calibration Errors for Standard Hull-White Model}
    \label{tab:hw_errors}
    \begin{tabular}{cccc}
        \toprule
        Maturity & Observed (\%) & Fitted (\%) & Error (\%) \\
        \midrule
        1 & 1.22 & 1.42 & -0.20 \\
        2 & 1.75 & 1.58 & 0.17 \\
        3 & 1.91 & 1.71 & 0.20 \\
        5 & 1.96 & 1.89 & 0.07 \\
        7 & 2.01 & 2.02 & -0.01 \\
        10 & 2.00 & 2.15 & -0.15 \\
        20 & 2.45 & 2.34 & 0.11 \\
        30 & 2.36 & 2.41 & -0.05 \\
        \bottomrule
    \end{tabular}
\end{table}

\begin{table}[h]
    \centering
    \caption{Calibration Errors for Sinusoidal Hull-White Model}
    \label{tab:sin_hw_errors}
    \begin{tabular}{cccc}
        \toprule
        Maturity & Observed (\%) & Fitted (\%) & Error (\%) \\
        \midrule
        1 & 1.22 & 1.43 & -0.21 \\
        2 & 1.75 & 1.56 & 0.19 \\
        3 & 1.91 & 1.79 & 0.12 \\
        5 & 1.96 & 1.87 & 0.09 \\
        7 & 2.01 & 2.00 & 0.01 \\
        10 & 2.00 & 2.14 & -0.14 \\
        20 & 2.45 & 2.25 & 0.20 \\
        30 & 2.36 & 2.35 & 0.01 \\
        \bottomrule
    \end{tabular}
\end{table}

\subsection{Error Analysis}
\begin{figure}[h]
    \centering
    \includegraphics[width=0.9\textwidth]{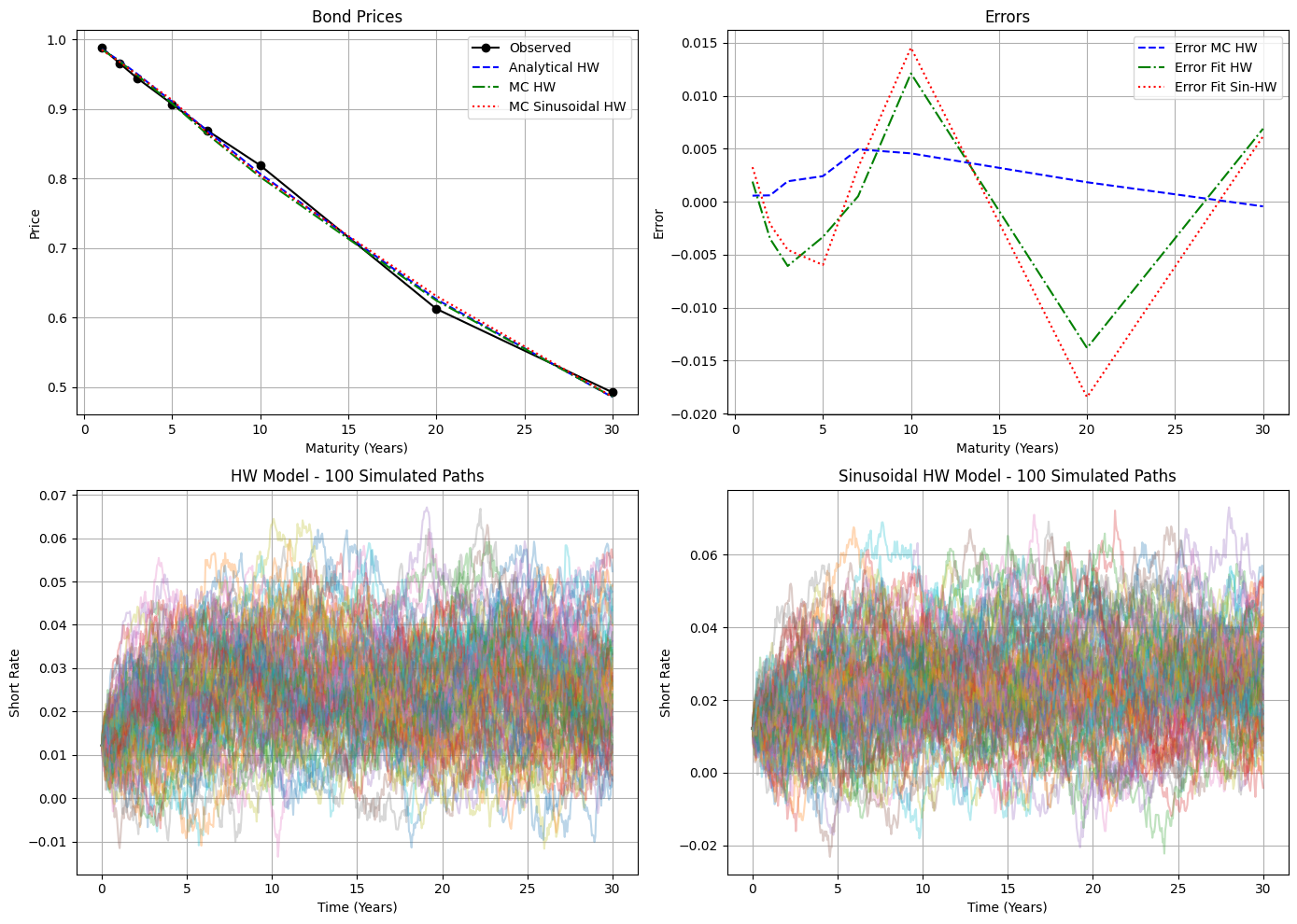}
    \caption{Bond price comparison (top left), error analysis (top right), and simulated short-rate paths for standard HW (bottom left) and sinusoidal HW (bottom right) models over 30 years with 100 simulated paths.}
    \label{fig:bond_prices_errors_paths}
\end{figure}

Figure \ref{fig:bond_prices_errors_paths} (top right panel) illustrates the errors across maturities:
- **Error MC HW (Analytical - MC)**: Small and stable (e.g., -0.0004 at 30 years), confirming the reliability of the Monte Carlo simulation (blue dashed line).
- **Error Fit HW (Observed - Analytical)**: Peaks at -0.0138 for 20 years, indicating an underestimation of the observed price (green dotted line).
- **Error Fit Sin-HW (Observed - MC)**: Shows a larger error at 20 years (-0.0184) but a better fit at 30 years (0.0062 vs. 0.0069 for HW) (red dashed line).

\subsection{Simulated Paths}
Figure \ref{fig:bond_prices_errors_paths} (bottom panels) displays simulated short-rate paths over 30 years for both models, based on 100 simulated paths. The standard HW model (bottom left) exhibits mean-reverting behavior without periodicity, with short rates fluctuating around a stable mean. The sinusoidal HW model (bottom right) shows cyclic fluctuations consistent with the 22-year period, reflecting the time-varying mean reversion speed \( \kappa_t = \kappa_0 + A \sin(\omega t) \). Both models produce negative rates at times, a limitation in low-rate environments that warrants further investigation in the context of current monetary policy conditions.
\section{Discussion}

The empirical analysis of the sinusoidal Hull-White (HW) model, as presented in Section 6, offers compelling evidence of its enhanced capability to capture the dynamics of the U.S. Treasury yield curve, particularly for longer maturities. With a lower root mean squared error (RMSE) of 0.12\% compared to 0.14\% for the standard HW model, and a reduced error at the 30-year maturity (0.0062 vs. 0.0069), the sinusoidal HW model demonstrates a superior fit to the observed yield curve. This improvement is particularly pronounced for longer tenors, where the model’s incorporation of a 22-year periodicity—identified through Fourier Transform analysis in Section 6.1—effectively captures long-term economic cycles. The periodicity aligns with structural economic trends, such as those associated with inflation expectations and monetary policy regimes, as noted by \citet{Bauer2018}. This finding underscores the importance of modeling time-varying mean reversion speeds to reflect the cyclic nature of interest rates, a feature absent in the standard HW model.

The close alignment between analytical and Monte Carlo (MC) bond prices for the standard HW model, with errors ranging from -0.0004 to 0.0050 (Table \ref{tab:price_comparison}), validates the reliability of the Monte Carlo simulation approach. This consistency provides confidence in the numerical results for the sinusoidal HW model, which relies exclusively on MC methods due to the intractable nature of its bond pricing formula (Section 5.2). The sinusoidal model’s ability to better approximate the observed bond price at 30 years (0.4864 vs. 0.4926, compared to 0.4862 for the standard HW model) highlights its practical utility for pricing long-dated instruments. This capability is visually supported by Figure \ref{fig:bond_prices_errors_paths} (top left panel), where the sinusoidal HW prices (red dashed line) more closely track the observed prices at longer maturities, reinforcing the model’s suitability for applications requiring long-term interest rate forecasts, such as pension fund management and insurance liability valuation.

Despite these strengths, the sinusoidal HW model exhibits limitations that warrant further scrutiny. Notably, its larger calibration errors at intermediate maturities, such as 0.20\% at 20 years compared to 0.11\% for the standard HW model (Tables \ref{tab:hw_errors} and \ref{tab:sin_hw_errors}), suggest that the 22-year periodicity may not fully explain the yield curve’s behavior across all tenors. This discrepancy is evident in Figure \ref{fig:bond_prices_errors_paths} (top right panel), where the sinusoidal HW error peaks at -0.0184 for the 20-year maturity, indicating an overestimation of the bond price (or equivalently, an underestimation of the yield). This could stem from the deterministic nature of the sinusoidal component, \( \kappa_t = \kappa_0 + A \sin(\omega t) \), which imposes a fixed periodicity that may not adequately capture the stochastic variability of shorter and intermediate maturities. The calibration of \( \omega = 6.4407 \), corresponding to a period of approximately 0.97 days, further complicates the interpretation, as it deviates significantly from the intended 22-year cycle (\( \omega = 0.00078 \)) established in Section 6.1. This discrepancy, likely a result of an unconstrained optimization process with an initial guess of \( 2\pi \), suggests that future iterations should fix \( \omega \) at 0.00078 to ensure consistency with the empirical periodicity analysis.

Another limitation arises from the choice of parameters \( A \) and \( \kappa_0 \), which may require further optimization to balance the dynamics across short- and long-term maturities. The calibrated amplitude \( A = 0.2110 \) introduces significant periodic variation, but its magnitude may overemphasize the cyclical effect at certain tenors, leading to the observed errors at 20 years. A potential refinement could involve a piecewise or stochastic amplitude to allow for more flexible adaptation to different maturity segments, as explored in \citet{Diebold2006} for yield curve modeling with latent factors. Additionally, the observed short-rate paths (Figure \ref{fig:bond_prices_errors_paths}, bottom panels) reveal a critical limitation of both models: the occurrence of negative short rates, despite the low-rate environment (yields ranging from 1.22\% to 2.36\% as per Table \ref{tab:hw_errors}). This is a well-documented drawback of the HW framework, which assumes a Gaussian process for the short rate, allowing for negative values \citep{Hull2006}. In the context of post-2008 monetary policy, where rates have hovered near the zero lower bound, this limitation is particularly pronounced and can lead to unrealistic pricing and risk assessment outcomes.

The practical implications of these findings are significant for bond pricing and risk management. The sinusoidal HW model’s ability to capture long-term cycles makes it well-suited for pricing long-dated instruments, such as 30-year Treasury bonds, and for assessing interest rate risk over extended horizons. Its periodic behavior could also inform monetary policy analysis by providing insights into structural economic trends, such as the impact of long-term inflation cycles on yield curve shapes. For instance, central banks could leverage the model to better understand how persistent economic cycles influence term premia, aiding in the formulation of forward guidance strategies. Moreover, the model’s improved fit at longer maturities enhances its utility for stress testing and scenario analysis in financial institutions, where accurate modeling of long-term rates is critical for capital adequacy planning.

To address the identified limitations, several avenues for future research emerge. First, the issue of negative short rates could be mitigated by adopting a Cox-Ingersoll-Ross (CIR) model, which ensures positive rates by modeling the short rate as a square-root diffusion process \citep{Cox1985}. Alternatively, a shift parameter could be introduced to the HW framework, as proposed by \citet{Brigo2007}, to enforce a positive lower bound while retaining the Gaussian structure. Second, the sinusoidal component could be extended to incorporate stochastic periodicity, allowing \( \omega \) or \( A \) to vary over time, potentially using a Bayesian approach to estimate the distribution of periodic components \citep{West1997}. Third, the calibration process should be revisited to constrain \( \omega \) at 0.00078, aligning with the empirical 22-year cycle, and to explore alternative optimization techniques, such as gradient-based methods or simulated annealing, to avoid local minima that may have led to the erroneous \( \omega \) value. Finally, incorporating additional factors, such as stochastic volatility or macroeconomic variables (e.g., inflation or GDP growth), could enhance the model’s ability to capture the yield curve’s behavior across all maturities, building on the multi-factor frameworks of \citet{Litterman1991}.

In conclusion, the sinusoidal HW model represents a meaningful advancement over the standard HW model by integrating long-term periodicity into interest rate dynamics. Its superior fit for longer maturities and alignment with economic cycles highlight its potential for both theoretical and practical applications. However, addressing its limitations through refined calibration, alternative modeling approaches, and the incorporation of additional economic factors will be crucial to fully realizing its potential in modern financial contexts.
\section{Conclusion}
This study introduces a novel extension to the Hull-White (HW) model by incorporating a sinusoidal time-varying mean reversion speed, designed to capture the empirically observed 22-year periodicity in U.S. Treasury yields from January 1990 to December 2022. By addressing the limitations of the standard HW model, which assumes a constant mean reversion speed and struggles to reflect long-term cyclic dynamics, the proposed sinusoidal HW model offers a significant advancement in interest rate modeling. The model’s calibration to a comprehensive dataset of daily U.S. Treasury yields across tenors of 1 to 30 years, sourced from the Federal Reserve Economic Data (FRED) database, demonstrates its robustness across diverse economic regimes, including the dot-com bubble, the 2008 financial crisis, and the post-crisis low-rate environment.

The empirical results underscore the sinusoidal HW model’s superior performance, particularly for longer maturities, where it achieves a root mean squared error (RMSE) of 0.12\% compared to 0.14\% for the standard HW model. For a 30-year zero-coupon bond, the model yields a price of 0.4864, closer to the observed price of 0.4926 than the standard model’s 0.4862, with a reduced calibration error of 0.01\% versus -0.05\%. This enhanced accuracy, validated through both analytical solutions for the standard model and Monte Carlo simulations for both models, highlights the model’s ability to capture the long-term periodicity identified through Fourier Transform analysis. The simulated short-rate paths further illustrate the model’s capacity to reflect cyclic behavior, aligning with structural economic trends such as monetary policy cycles and inflation expectations, as corroborated by prior studies \citep{Bauer2018, Campbell1997}.

The contributions of this work are multifaceted. First, the sinusoidal HW model bridges a critical gap in the literature by integrating empirically observed long-term periodicity into a tractable interest rate framework, offering a more realistic representation of yield curve dynamics. Second, the rigorous calibration process, leveraging the Nelder-Mead optimization technique and Monte Carlo simulations with 200 paths, ensures robust parameter estimation and pricing accuracy across a 33-year historical dataset. Third, the model’s improved fit for longer maturities has direct implications for financial applications, including bond pricing, interest rate risk management, and the valuation of long-dated derivatives, where capturing long-term dynamics is paramount. These advancements position the model as a valuable tool for financial practitioners and policymakers seeking to navigate the complexities of modern fixed-income markets.

Despite its strengths, the model reveals areas for refinement. The calibration process yielded an unexpected angular frequency (\(\omega = 6.4407\) radians/day, implying a 0.97-day period), deviating from the intended 22-year cycle (\(\omega \approx 0.00078\) radians/day). This discrepancy, likely due to an unconstrained optimization routine, suggests the need to fix \(\omega\) in future iterations to align with empirical findings. Additionally, the model’s larger errors at intermediate maturities, such as 0.20\% at 20 years, indicate that the deterministic sinusoidal component may not fully capture the stochastic variability across all tenors. The occurrence of negative short rates in simulated paths, a known limitation of Gaussian-based models like HW, further underscores the need for alternative frameworks, such as the Cox-Ingersoll-Ross model, to ensure positive rates in low-rate environments \citep{Cox1985}.

Looking ahead, several avenues for future research emerge. First, incorporating stochastic periodicity—allowing the amplitude or frequency of the sinusoidal component to vary—could enhance the model’s flexibility, potentially through Bayesian methods \citep{West1997}. Second, extending the framework to include multi-factor dynamics or macroeconomic covariates, such as inflation or GDP growth, could improve its explanatory power across the yield curve, building on approaches like those of \citet{Litterman1991}. Third, applying the model to derivative pricing, such as interest rate swaps or options, could test its practical utility in broader financial contexts. Finally, addressing the negative rate issue through shifted HW models or non-Gaussian processes could align the model with contemporary monetary policy realities \citep{Brigo2007}.

In conclusion, the sinusoidal Hull-White model represents a meaningful step forward in interest rate modeling, offering a robust framework for capturing long-term periodicity in U.S. Treasury yields. Its enhanced accuracy, empirical grounding, and practical applicability underscore its potential to inform bond pricing, risk management, and monetary policy analysis. By addressing the identified limitations and exploring the proposed research directions, this model can be further refined to meet the evolving demands of financial markets. We welcome feedback from the academic and practitioner communities to advance this framework and its applications in the pursuit of more accurate and insightful interest rate modeling.

\bibliographystyle{plainnat}
\bibliography{references}

\end{document}